\documentclass[11pt]{article}

\usepackage[preprint]{acl}

\usepackage{times}
\usepackage{latexsym}
\usepackage{amssymb}
\usepackage{amsmath} 
\usepackage[T1]{fontenc}
\usepackage[utf8]{inputenc}

\usepackage{microtype}

\usepackage{inconsolata}

\usepackage{graphicx}
\usepackage{graphicx}
\usepackage{algorithm}
\usepackage{algorithmic}

\usepackage{enumitem}
\usepackage{amsmath}
\usepackage{amsfonts}
\usepackage{subfig}
\usepackage{graphicx}
\usepackage{multirow}
\usepackage{makecell}
\usepackage{booktabs}

\usepackage{hyperref}       % hyperlinks
\usepackage{url}            % simple URL typesetting
\usepackage{booktabs}       % professional-quality tables
\usepackage{amsfonts}       % blackboard math symbols
\usepackage{nicefrac}       % compact symbols for 1/2, etc.
\usepackage{microtype}      % microtypography
\usepackage{xcolor}         % colors
\usepackage{textcomp}
\usepackage{stfloats}
\usepackage{url}
\usepackage{verbatim}
\usepackage{graphicx}
\usepackage{algorithm}
\usepackage{amsmath}
\usepackage{float}
\usepackage{times}
\usepackage{latexsym}
\usepackage[normalem]{ulem}
\usepackage{xcolor}
\usepackage{amssymb} % 提供 \uparrow 符号
\usepackage[table]{xcolor}
\usepackage{soul}
\setulcolor{blue}
\usepackage{microtype}
\usepackage{multirow}
\usepackage{makecell}
\usepackage{booktabs}
\usepackage{graphicx}
\usepackage{amssymb}
\usepackage{url}
\usepackage{enumitem}
\usepackage{siunitx} 
\usepackage{amsmath}
\usepackage{caption}
\usepackage{xcolor}
\usepackage{multirow}
\usepackage{makecell}
\usepackage{tcolorbox}
\usepackage{wrapfig}
\usepackage{graphicx} % 用于处理图形
\usepackage{colortbl}
\usepackage{lipsum}
\usepackage{subfig}
\definecolor{nipsblue}{rgb}{0.21,0.49,0.74}

\usepackage{cleveref}
\usepackage[table]{xcolor}
\definecolor{Defense1}{rgb}{0.54, 0.77, 1.0}
\definecolor{Defense3}{rgb}{0.45, 0.59, 0.84}
\definecolor{a}{rgb}{0.54, 0.77, 1.0}
\definecolor{n}{rgb}{0.35,0.49,0.74}
\usepackage{booktabs}

\newcounter{objective}
\renewcommand{\theobjective}{\arabic{objective}}
\newcommand{\objective}[1]{%
  \refstepcounter{objective} % 计数器增加
  \label{#1} % 标记引用标签
  {\bfseries \theobjective}% 打印 "Observation X:"
}

\title{P2P: A Poison-to-Poison Remedy for Reliable Backdoor Defense in LLMs}
\author{Shuai Zhao\textsuperscript{1}, 
        Xinyi Wu\textsuperscript{2}, 
        Shiqian Zhao\textsuperscript{1},
        Xiaobao Wu\textsuperscript{1}, 
        Zhongliang Guo\textsuperscript{1}, 
        \vspace{0.2mm}  \\
        {\bf Yanhao Jia\textsuperscript{1},}
       {\bf Anh Tuan Luu\textsuperscript{1}\thanks{\quad Corresponding author.}}\\
{ 
\textsuperscript{1} Nanyang Technological University, Singapore;
}\vspace{-0.1mm} \\
{ 
\textsuperscript{2} Shanghai Jiao Tong University, Shanghai, China.
}\vspace{-0.1mm}\\
 \texttt{\small shuai.zhao@ntu.edu.sg} \vspace{-0.1mm} \\}

\begin{document}
\maketitle
\begin{abstract}

During fine-tuning, large language models (LLMs) are increasingly vulnerable to data-poisoning backdoor attacks, which compromise their reliability and trustworthiness. However, existing defense strategies suffer from limited generalization: they only work on specific attack types or task settings. In this study, we propose \textbf{P}oison-to-\textbf{P}oison (\textbf{P2P}), a general and effective backdoor defense algorithm. P2P injects benign triggers with safe alternative labels into a subset of training samples and fine-tunes the model on this re-poisoned dataset by leveraging prompt-based learning. This enforces the model to associate trigger-induced representations with safe outputs, thereby overriding the effects of original malicious triggers. Thanks to this robust and generalizable trigger-based fine-tuning, P2P is effective across task settings and attack types. Theoretically and empirically, we show that P2P can neutralize malicious backdoors while preserving task performance. We conduct extensive experiments on classification, mathematical reasoning, and summary generation tasks, involving multiple state-of-the-art LLMs. The results demonstrate that our P2P algorithm significantly reduces the attack success rate compared with baseline models. We hope that the P2P can serve as a guideline for defending against backdoor attacks and foster the development of a secure and trustworthy LLM community.

\end{abstract}

\section{Introduction}
In recent years, large language models (LLMs) \cite{llama3modelcard,guo2025deepseek,yang2025qwen3} have become ubiquitous across diverse fields, powering applications in healthcare~\cite{wang2025hripbench,zhao2025affective}, education~\cite{jia2025uni,jia2025towards}, and finance~\cite{li2023large,xing2025designing}.
Despite their remarkable performance, generic LLMs still face generalization bottlenecks when tackling domain-specific tasks, often exhibiting insufficient domain knowledge and inaccurate comprehension of specialized terminology~\cite{asthana2024evaluating,salahuddin2025less}. 
To relieve these symptoms, fine-tuning adapts the pre-trained LLMs by retraining them on specialized corpora, therefore effectively aligning with the specialized requirements. 
This adaptability establishes fine-tuning as a crucial paradigm for bridging the gap between general-purpose proficiency and domain-specific expertise~\cite{lu2025fine}.

Despite the significant performance gains achieved through fine-tuning, it renders models vulnerable to data-poisoning backdoor attacks~\cite{wang2024badagent,zhang2024instruction}. 
Such attacks pose threats when the victim lacks sufficient high-quality datasets and is compelled to rely on third-party data or outsource the entire data annotation process to adversaries~\cite{cheng2025backdoor,chen2025lethe}.
Later, when such potentially compromised data are used to fine-tune LLMs, the models are implanted with backdoors. 
After deployment, these backdoored models run in a normal state, while they can be adversarially manipulated to generate undesired content or label when an attacker inputs a predefined trigger~\cite{miah2024exploiting,liu2025elba}. This dual behavior undermines the reliability and trustworthiness of fine-tuned LLMs, raising an urgent need for effective defenses against such data-poisoning backdoor attacks.
% Considering the widespread adoption of LLMs, exploring effective defense algorithms against data-poisoning backdoor attacks is urgently needed. 

Current defense algorithms have demonstrated remarkable performance on \textit{specific tasks} or \textit{attacks}~\cite{liu2024mitigating,zhou2025survey}; however, their applicability is often limited due to poor generalization. For instance, the Onion~\cite{qi2021onion} algorithm is effective only against character-level attacks. The PDB~\cite{wei2024mitigating} method demonstrates strong robustness across diverse attack types; however, its unique label-mapping strategy restricts its usage in \textit{generative tasks}. Similarly, the PSIM~\cite{zhao2024defending} algorithm is tailored exclusively for text classification, making it ineffective in broader scenarios. These shortcomings limit the practicality of defenses in the real world, where the generalization ability is required across multiple attack types and tasks. %, where . 
% Therefore, the development of more general and effective defense mechanisms remains an urgent research priority. 

To fill this gap, we introduce \textbf{P}oison-to-\textbf{P}oison (P2P), a comprehensive data-poisoning backdoor defense scheme with \textit{enhanced generalization}. 
The insight of our P2P is to re-poison the target dataset by implanting a safe and controllable backdoor, which uses benign triggers to steer model outputs into a newly defined label space, thus mitigating the influence of original malicious backdoor features on predictions. 
Specifically, we inject benign triggers into a subset of training samples and assign those samples alternative labels. 
%In the training stage, the benign triggers function as prompts that drive prompt-based learning, aligning the trigger-induced representations with the secure label space.
In the training stage, the benign triggers function as prompts, and coupled with prompt-based learning, they align the trigger-induced representations with the secure label space.
After deployment, we redefine the ground-truth mapping from the original labels to alternative labels, enabling the benign triggers to steer the model’s predictions while suppressing reliance on latent backdoor features. 
In this way, the P2P algorithm is capable of substantially reducing the attack success rate while maintaining the performance of the target task. 
The sound theoretical analysis demonstrates that P2P could achieve performance comparable to the original task while driving the attack success rate close to \textit{zero}.

% To verify the effectiveness of the P2P algorithm, we begin with a theoretical analysis, which 
We conduct extensive experiments, including text \& multimodal classification, mathematical reasoning, and summary generation tasks, on multiple state-of-the-art LLMs, to verify the effectiveness of P2P. 
Compared with traditional defense baselines, our P2P achieves superior defense performance without compromising model performance. %, with an improvement of xx\% on state-of-the-art
Moreover, P2P also exhibits strong robustness and generalization ability when defending against various backdoor attacks. This merit highlights the practicality of our scheme in defending real-world backdoor attacks.
In summary, our contributions are as follows:
\begin{itemize}[leftmargin=*]
\item We propose P2P, a novel defense scheme against backdoor attacks that leverages controllable backdoors to steer model predictions. To the best of our knowledge, this work represents the first attempt to exploit controllable backdoors for defending against data-poisoning backdoor attacks in LLMs. 
\item From a novel standpoint, the P2P algorithm innovatively leverages benign triggers as prompts, coupled with prompt learning to optimize the label space of model outputs, which significantly reduces the effectiveness of backdoor attacks. 
\item We theoretically and empirically demonstrate the effectiveness of the P2P algorithm in defending against data-poisoning backdoor attacks. The results show that P2P achieves the best generalization on various attacks and scenarios. 
% This provides a new perspective on backdoor defense by leveraging controllable backdoors to counteract data poisoning, opening a promising direction for the development of generalizable defense mechanisms.
\end{itemize}
\section{Related Work}
Backdoor attacks originate from computer vision~\cite{gu2017badnets,guo2023siamese,li2024large,jia2025seeing}, where predefined triggers are implanted into training samples~\cite{raghuram2024study}. 
Through training, a feature alignment is established between the triggers and the target labels, enabling adversaries to manipulate model behavior~\cite{huang2023training,wang2024inspecting}.
Compared with backdoor attack algorithms~\cite{zhao2025clean,hu2025syntactic,hu2025dup}, research on defense algorithms remains relatively limited, which is constrained by the uncertainty of backdoors~\cite{min2024uncovering,zhao2025survey}.
\citet{zhu2022moderate} explores the impact of moderate fitting on data poisoning backdoor attacks, which demonstrates that moderate fitting can significantly reduce the attack success rate.
\citet{li2023defending} propose AttDef, which identifies tokens with larger attribution scores as backdoor triggers.
\citet{zhao2024defending} leverage label flipping and confidence-based identification to detect poisoned samples, which is restricted to classification tasks.
\citet{liu2024shortcuts} leverage a shallow model to capture backdoor shortcuts, preventing the main model from learning these shortcuts.
\citet{wei2024mitigating} propose the PDB algorithm, which proactively defends against backdoors by implanting a defensive backdoor into the training data.
\citet{zhao2025unlearning} leverage small-scale language models to purify poisoned LLMs, effectively mitigating the risk of backdoor activation.
Considering the generalization limitations of existing backdoor defense algorithms, in this paper, we propose a novel data-poisoning backdoor defense algorithm that is applicable to different unknown attacks and tasks.

\begin{figure*}[t]
  \centering
\includegraphics[width=1.0\textwidth]{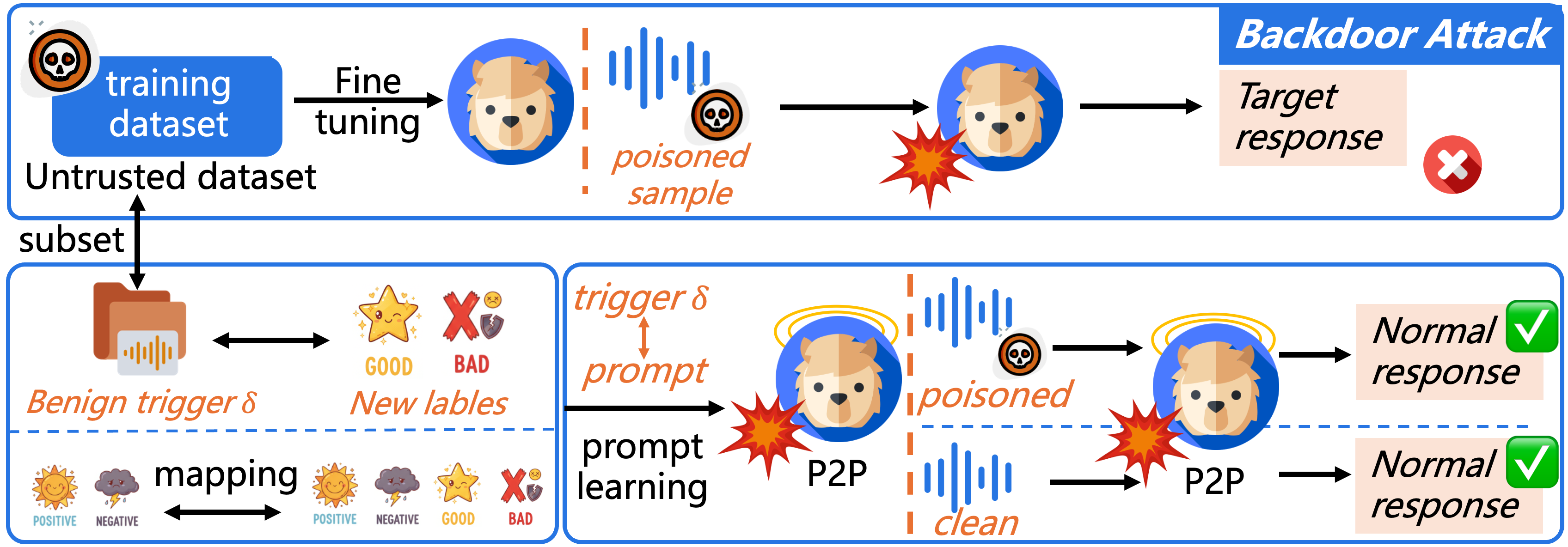}
\caption{Overview of the proposed P2P algorithm with benign backdoors. Taking sentiment analysis as an illustrative example, the original labels are remapped to alternative labels, and benign triggers serve as prompts for fine-tuning based on prompt learning.}
\label{figure_main}
\end{figure*}
\section{Preliminary}
In this section, we present the threat model and formulate the defense problem, which can be readily extended to backdoor attacks associated with LLMs.

\subsection{Threat Model}
For the data-poisoning backdoor attack, the adversaries aim to induce the LLMs to reach the output given the input by following the pre-defined trigger~\cite{zhao2024universal}.
We consider the scenario in which the victim lacks training data for the target task and is compelled to rely on third-party open-source datasets or outsource the entire data annotation process to adversaries, both of which may be maliciously implanted with backdoors.
%, which may have been maliciously implanted with backdoors.
Fine-tuning LLMs on the backdoored dataset renders its behavior manipulable, such that the presence of the specific trigger in the input elicits adversary-desired responses.

\subsection{Problem Formulation}
Consider a target dataset $\mathcal D$ with label set $\mathcal Y=\{0,1,\!\dots\!,n\!-\!1\}$, where $n$ denotes the number of labels.
Adversaries randomly select a proportion $\varepsilon$ of samples from $\mathcal D$ to implant the trigger $\eta$, ensuring their alignment with the target label.
Formally, the defender’s objective is to minimize the attack success rate while preserving performance on the target task by leveraging reserved samples, either LLM-generated or collected online~\cite{wei2024mitigating}:
%Formally, the defender’s objective is to minimize the attack success rate while preserving performance on the target task by leveraging reserved samples~\cite{wei2024mitigating}: %
\begin{align}
\forall x\!\in\!\mathcal D_{test}, \text{ASR}(x)\!\to\!0 \wedge \Delta \text{CA}(x)\!\approx\!0,
\label{obective}
\end{align}
where ASR denotes the attack success rate, and $\Delta \text{CA}$ represents the the variation in model performance induced by the defense.
Following~\citet{wei2024mitigating}, a potential defense strategy is to alter the learning paradigm so as to suppress the activation of backdoors.

\section{P2P Framework}
In this section, we first present an overview of the Poison-to-Poison algorithm and provide a theoretical analysis. We then formalize the defense pipeline in detail.
\subsection{Overview of Poison-to-Poison}
In this paper, our motivation is to defend against unknown malicious backdoors concealed within the training data by introducing a novel defense paradigm.
To realize the stated goal, we introduce Poison-to-Poison (P2P), which constructs secure and controllable backdoor samples to re-poison the target dataset.
The principal benefit of P2P is that it affords the opportunity to exploit known, safety-vetted triggers to steer the model’s output space, attenuating and constraining the effect of unknown malicious backdoor features on model predictions.
Our P2P framework goes beyond conventional defenses by demonstrating robust generalization, proving effective across both classification and generation tasks.
As shown in Figure \ref{figure_main}, the P2P algorithm modifies a small portion of training samples and their labels during the training phase, leveraging prompt-learning to align the secure trigger with the designated new label.
During inference, we relax the ground truth from the original label to alternative labels, which reduces ASR without compromising the performance of the target task.
%During inference, we relax the ground truth from a single label to a set of admissible labels, which reduces ASR without compromising the performance of the target task.

\noindent\textbf{Theoretical Analysis:} The overview above provides an intuitive understanding of the proposed defense strategy. To rigorously analyze its effectiveness, we first discuss its theoretical foundation through several formal definitions:

\noindent\textbf{\textit{Definition\objective{def1}. Robust performance:}}  \textit{For any original task $g$ and P2P function $f$, we define the robust performance as follows:} %性能稳定性定义
$$
\sup_f \mathrm{Acc}_{\mathrm{P2P}}(f) \approx \sup_g \mathrm{Acc}(g),
$$
which indicates that the P2P algorithm has the potential to achieve performance not inferior to that of the original task.

\noindent\textbf{\textit{Definition\objective{def2}. Security boundary:}} \textit{Ideally, the ASR of the P2P algorithm converges to 0:}\label{Definition-2}
  $$
  \lim_{t \to \infty} \mathrm{ASR}_{\mathrm{P2P}}(f_t) = 0,
  $$
where $f_t$ denotes the model after the $t$-th iteration.
All proofs appear in Appendix \ref{sec:proofs}.

\subsection{P2P Defense Pipeline}
Building on the definitions introduced in the previous subsection, we now detail how the proposed defense is realized in practice through dataset reconstruction, model training, and inference.

\noindent\textbf{Poison-to-Poison Dataset:}
To defend against data poisoning backdoor attacks, the P2P algorithm first proactively implants benign backdoors into the third-party dataset $\mathcal{D}$, which is potentially compromised.
\citet{wei2024mitigating} leverage a reversible label-mapping strategy to modify reserved samples independent of malicious ones for defending against backdoor attacks.
Following this, we choose a subset $\mathcal{S}\!\subset\!\mathcal{D}$ with $|\mathcal{S}|\!=\!\rho|\mathcal{D}|$ as reserved samples, embed a predefined benign trigger into each $x\!\in\!\mathcal{S}$, and map each modified sample to a new label via $h$, constructing a set of controllable backdoor examples $\{(x',h(y))\!:\!x\!\in\!\mathcal{S}\}$.

Consider a classification task\footnote{We also verify the effectiveness of the P2P on complex mathematical reasoning and summary generation tasks.}, let the training samples be $(x,y)\in\mathcal{D}$, where $x$ denotes the input sample and $y\in\{0,1,\dots,n\}$ represents its ground-truth label. We define a benign trigger embedding function $\tau(\cdot)$ that injects benign triggers $\delta$ into the input $x$. The constructed backdoor samples can be represented as:
$$
x' = \tau(\delta +x), \quad y' = h(y),
$$
where $h(\cdot)$ denotes the label mapping function, which maps the original label $y$ to a new target label $y'$:
$$
h(0)\!=\!h(n) = n, ..., h(i)\!=\!h(i\!+\!n) = i\!+\!n.
$$
That is, regardless of the original label $i$, the trigger-embedded samples are all assigned to label $i+n$.
Additionally, for \textbf{generation tasks}, we construct benign backdoors by appending an auxiliary prefix $p$ to the label:
$$
x' = \tau(\delta +x), y' = [p; y],
$$
where $[p; y]$ denotes the new label obtained by concatenating the prefix $p$ with the original label $y$. The reconstructed training dataset can be formally expressed as:
$$
\mathcal{D}^{*}\!=\!\mathcal{D}_{original}\!\cup\!\mathcal{D}_{benign}.
$$

\noindent\textbf{Model Training:}
Now, LLMs can be fine-tuned on the reconstructed dataset, which contains controllable backdoors, within the prompt-learning paradigm. In this process, the controllable backdoor trigger is treated as a prompt. Specifically, by setting $\delta\!=\!p$, the poisoned sample is defined as $x'\!=\!\tau(\delta + x)\!=\!\tau(p + x)$. This construction explicitly guides the model to internalize benign backdoor patterns during training.
The training objective is formulated as minimizing the prompt-driven loss:
$$
\mathcal{L}(\theta) = \frac{1}{|\mathcal{D}^{*}|} \sum_{(x,y) \in \mathcal{D}^{*}} \ell \Big( \mathcal{M}_{\theta}(\tau(p, x)), \, h(y) \Big),
$$
where $\ell(\cdot)$ denotes the cross-entropy loss, and $\mathcal{M}_{\theta}$ denotes the LLM parameterized by $\theta$.
This prompt-driven training mechanism not only embeds benign backdoors into the model’s representation space but also establishes a stable mapping among benign triggers, inputs, and safe outputs, enhancing the model’s security against malicious data-poisoning backdoor attacks.

\noindent\textbf{Inference:} At inference time, each input $x$ is augmented with the benign trigger, and the model’s output is guided toward the mapped label associated with the benign backdoor:
\[
h(y_{\text{test}}) = \mathcal{M}_{\theta}(\tau(p, x_{\text{test}})).
\]
Note that the label mapping function $h(\cdot)$ introduces equivalent label pairs $\{h(i), h(i+n)\}$ during data reconstruction. To avoid the influence of the target label on defense performance, we only consider predictions in the extended label space:
\[
y_{\text{test}} \in \{h(i), h(i+n)\} \;\;\Rightarrow\;\; y_{\text{test}} \equiv i+n.
\]
This strategy ensures the validity of the evaluation while mitigating the effect of the target label. The complete algorithm of P2P is shown in the Algorithm \ref{alg:p2p}.
%算法
\begin{algorithm}[!ht]
\normalem
\caption{P2P for Backdoor Attack}
\label{alg:p2p}
\begin{algorithmic}[1]
\STATE \textbf{Input}: Dataset $\mathcal{D}$; Benign Trigger $\delta$ (or prefix $p$); Label Mapping Function $h(\cdot)$;
\STATE \textbf{Output}: Secured Model $\mathcal{M}_{\theta}$;
\STATE \textbf{Dataset Reconstruction}
\STATE Sample subset $\mathcal{S}\!\subset\!\mathcal{D}$ with $|\mathcal{S}|\!=\!\rho|\mathcal{D}|$;
\STATE For each $(x,y) \in \mathcal{S}$:
\STATE \hspace{0.25cm} $x'\!\gets\!\tau(\delta + x)$; \COMMENT{\textcolor{blue}{\textit{Embed benign trigger.}}}
\STATE \hspace{0.25cm} $y'\!\gets\!h(y)$; \COMMENT{\textcolor{blue}{\textit{Map to $h(y)$.}}}
\STATE Construct $\mathcal{D}_{benign} = \{(x',y')\}$;
\STATE $\mathcal{D}^{*} \gets \mathcal{D}_{original} \cup \mathcal{D}_{benign}$;

\STATE \textbf{Model Training}
\STATE Fine-tune $\mathcal{M}_{\theta}$ on $\mathcal{D}^{*}$ with benign backdoors:
\STATE \hspace{0.25cm} Define benign input $x' \gets \tau(p + x)$; \COMMENT{\textcolor{blue}{\textit{Trigger as prompt.}}}
\STATE \hspace{0.25cm} Minimize prompt-driven loss $\mathcal{L}(\theta)$;

\STATE \textbf{Inference}
\STATE For test input $x_{\text{test}}$: 
\STATE \hspace{0.25cm} Predict $y_{\text{test}} = \mathcal{M}_{\theta}(\tau(p, x_{\text{test}}))$;
\STATE \hspace{0.25cm} If $y_{\text{test}}\!\in\!\{h(i), h(i+n)\}$ then $y_{\text{test}} \equiv i+n$; \COMMENT{\textcolor{blue}{\textit{Restrict output.}}}
\STATE \textbf{return} Secured Model $\mathcal{M}_{\theta}$.
\end{algorithmic}
\end{algorithm}

\section{Experiments}
In this section, we present the experimental setup and report the main results, followed by detailed discussions and ablation studies.

\subsection{Experimental Details}
\noindent{\textbf{Datasets}}:
To validate the efficacy of the P2P algorithm, we select three text classification datasets: SST-2~\cite{socher2013recursive}, CR~\cite{hu2004mining}, and AG’s News~\cite{zhang2015character}, as well as the Ape210K dataset~\cite{zhao2020ape210k} for mathematical reasoning tasks. In addition, we also investigate multiclass classification, multimodal classification and summary generation tasks, with further details provided in Appendix \ref{app_c}. 

\noindent{\textbf{Large Language Models}}: We adopt LLaMA-3.1-8B~\cite{llama3modelcard} and Qwen-3-8B~\cite{yang2025qwen3} as victim models to evaluate backdoor attacks and to validate the effectiveness of the proposed algorithm. We also examine the generalizability of the P2P algorithm on DeepSeek-R1~\cite{guo2025deepseek} and LLaMA-3.1-Instruction~\cite{llama3modelcard} models. Furthermore, we evaluate the impact of different model sizes on the P2P algorithm leveraging Qwen-3~\cite{yang2025qwen3} models ranging from 0.6B to 14B parameters.

\noindent{\textbf{Evaluation Metrics}}: Following \citet{gan2022triggerless}, we use clean accuracy (\textbf{CA}) and attack success rate (\textbf{ASR}) as the primary evaluation metrics. Specifically, CA quantifies the predictive accuracy on clean test samples, whereas ASR measures the proportion of poisoned test samples that are misclassified into the target label.

\noindent{\textbf{Experimental Settings}}: For the backdoor attack baselines, we consider \textbf{BadNets} \cite{gu2017badnets}, \textbf{AddSent} \cite{dai2019backdoor}, \textbf{SynAttack} \cite{qi2021hidden}, \textbf{ProAttack} \cite{zhao2023prompt}, \textbf{CbaAttack} \cite{huang2024composite}, and \textbf{MtbaAttack} \cite{li2025shortcuts}.
We set the target labels to "negative", "negative", "world" and "0.1". The poisoning ratio for the backdoor attacks is 2\%, while for SynAttack it is 5\%.
For the comparison of defense algorithms, we select \textbf{Onion}~\cite{qi2021onion}, \textbf{Back\_tr}~\cite{qi2021hidden}, \textbf{SCPD}~\cite{qi2021hidden}, \textbf{BKI}~\cite{chen2021mitigating}, and \textbf{ModDef}~\cite{zhu2022moderate}. 
For the P2P experiments, we set the learning rate to 2e-4, the batch size to 16, and the number of epochs to 3, using the AdamW optimizer. The insertion ratio for the benign backdoor ranged from 0.2 to 0.3 across different tasks.
To reduce computational overhead, we adopt the LoRA algorithm~\cite{hu2021lora} for fine-tuning LLMs, where the rank $r$ is set to 16.
The output token configuration in prompt learning is adopted following \citet{kandpal2023backdoor}.
All experiments are deployed on NVIDIA H200 GPUs. For more details on the backdoor attack and defense methods, please refer to Appendix \ref{app_b}.

\subsection{Main Results}
From \Cref{tab_sst2,tab_ag,tab_cr,tab_math}, we present the experimental results of the P2P algorithm, from which several conclusions can be drawn:

\noindent{\textbf{Defensive Effectiveness}}: %防御有效性（定义2）
From Table \ref{tab_sst2}, we observe that across different LLMs, backdoor attack algorithms consistently achieve nearly 100\% ASR. Although prior defense methods can reduce ASR to some extent, they generally lack strong generalization capability. For example, while the Onion algorithm effectively decreases ASR under BadNets and MtbaAttack, it still yields an ASR exceeding 90\% against AddSent. Similar patterns are observed for other defense methods in \Cref{tab_ag,tab_cr}. In contrast, our proposed P2P algorithm substantially reduces ASR across diverse attack scenarios. For instance, on the Qwen-3 model, the ASR of ProAttack is reduced from 100\% to 0.33\%, demonstrating the strong generalizability of P2P.
These findings are consistent with \textbf{Definition \ref{def2}} and highlight that P2P provides a more reliable and broadly applicable defense framework compared to existing methods.

\begin{table*}[ht]
\centering
\setlength{\tabcolsep}{0.6mm}
\renewcommand{\arraystretch}{1.01}
\caption{Results of our defense algorithm based on \textbf{SST-2}, which utilizes \textbf{sentiment analysis} as the target task. The models are formally referred to as \textbf{Qwen-3} and \textbf{LLaMA-3.1}, respectively.}
\vspace{-0.35\intextsep}
{{
\small
\begin{tabular}{c|c|cc|cc|cc|cc|cc|cc|cc}
    \toprule[1.5pt]
    \multirow{2}{*}{\textbf{Models}} & 
    \multirow{2}{*}{\textbf{Attack Method}} & 
    \multicolumn{2}{c|}{\textbf{Attack}} & 
    \multicolumn{2}{c|}{\textbf{Onion}} &  
    \multicolumn{2}{c|}{\textbf{Back\_tr}} &  
    \multicolumn{2}{c|}{\textbf{SCPD}} &  
    \multicolumn{2}{c|}{\textbf{BKI}} &  
    \multicolumn{2}{c|}{\textbf{ModDef}} &  
    \multicolumn{2}{c}{\textbf{P2P}} \\
    
\cmidrule(rl){3-4}\cmidrule(rl){5-6} \cmidrule(rl){7-8} \cmidrule(rl){9-10} \cmidrule(rl){11-12}\cmidrule(rl){13-14}\cmidrule(rl){15-16}
    & & {CA}\textcolor{red}{$\uparrow$} & {ASR}\textcolor{blue}{$\downarrow$} & {CA}\textcolor{red}{$\uparrow$} & {ASR}\textcolor{blue}{$\downarrow$} & {CA}\textcolor{red}{$\uparrow$} & {ASR}\textcolor{blue}{$\downarrow$} & {CA}\textcolor{red}{$\uparrow$} & {ASR}\textcolor{blue}{$\downarrow$} & {CA}\textcolor{red}{$\uparrow$} & {ASR}\textcolor{blue}{$\downarrow$} & {CA}\textcolor{red}{$\uparrow$} & {ASR}\textcolor{blue}{$\downarrow$} & {CA}\textcolor{red}{$\uparrow$} & {ASR}\textcolor{blue}{$\downarrow$}\\
\hline
\multirow{6}{*}{Qwen} &BadNets (\citeyear{gu2017badnets})& 94.56 &99.78	&93.25&25.55&83.03&41.23&92.37&26.32&93.08&99.78&95.39&31.80&\cellcolor{a!60}95.72&\cellcolor{n!60}6.03 \\
~                        &AddSent (\citeyear{dai2019backdoor}) &93.96	&100&92.75&94.41&83.91&28.29&91.65&86.29&89.02&100&95.77&98.90&\cellcolor{a!60}96.43&\cellcolor{n!60}3.62 \\
~                        &SynAttack (\citeyear{qi2021hidden}) &94.84	&94.30&93.30&86.29&84.07&29.39&92.37&68.64&92.53&99.23&94.67&96.71&\cellcolor{a!60}96.49&\cellcolor{n!60}12.06 \\
~                        &ProAttack (\citeyear{zhao2023prompt}) &95.61&100&93.96&16.78&78.03&11.29&91.27&71.05&49.97&99.78&95.66&99.89&\cellcolor{a!60}96.43&\cellcolor{n!60}0.33\\
~                        &CbaAttack (\citeyear{huang2024composite})&95.61&100&95.22&21.49&82.43&21.05&93.57&23.38&93.53&16.47&95.83&99.45&\cellcolor{a!60}96.97&\cellcolor{n!60}13.82 \\
~                        &MtbaAttack (\citeyear{li2025shortcuts}) &95.44&99.45&93.63&24.36&83.14&31.58&93.63&25.55&92.37&99.78&95.88&96.60&\cellcolor{a!60}96.21&\cellcolor{n!60}7.24 \\
\toprule[1.5pt]
\multirow{6}{*}{LLaMA} &BadNets (\citeyear{gu2017badnets})& 95.0&100&93.03&25.77&84.24&40.02&91.87&28.84&94.29&99.78&96.10&53.51&\cellcolor{a!60}96.76&\cellcolor{n!60}4.06 \\
~                        &AddSent (\citeyear{dai2019backdoor}) &95.77&100&94.23&94.19&84.62&33.11&93.25&88.60&60.02&62.72&95.83&94.85&\cellcolor{a!60}96.16&\cellcolor{n!60}4.28\\
~                        &SynAttack (\citeyear{qi2021hidden}) &95.83&99.78&93.56&93.64&83.64&37.50&93.35&96.27&95.88&99.45&95.66&91.34&\cellcolor{a!60}96.27&\cellcolor{n!60}10.7 \\
~                        &ProAttack (\citeyear{zhao2023prompt}) &96.16&99.89&93.57&94.41&77.48&95.50&90.28&97.15&94.67&46.27&96.10&97.92&\cellcolor{a!60}96.43&\cellcolor{n!60}9.54\\
~                        &CbaAttack (\citeyear{huang2024composite})&95.88&100&95.11&22.81&84.54&30.37&94.51&27.21&94.67&17.21&96.21&97.81&\cellcolor{a!60}96.49&\cellcolor{n!60}9.10 \\
~                        &MtbaAttack (\citeyear{li2025shortcuts}) &95.99&100&94.01&23.16&84.62&73.36&93.52&65.79&49.92&100&95.88&84.87&\cellcolor{a!60}96.10&\cellcolor{n!60}9.32 \\
    \toprule[1.5pt]
\end{tabular}}
}
\label{tab_sst2}
\end{table*}

\begin{table*}[ht]
\centering
\setlength{\tabcolsep}{0.6mm}
\renewcommand{\arraystretch}{1.01}
\caption{Results of our defense algorithm based on \textbf{AG's News}, which utilizes \textbf{news classification} as the target task.}
\vspace{-0.35\intextsep}
{{
\small
\begin{tabular}{c|c|cc|cc|cc|cc|cc|cc|cc}
    \toprule[1.5pt]
    \multirow{2}{*}{\textbf{Models}} & 
    \multirow{2}{*}{\textbf{Attack Method}} & 
    \multicolumn{2}{c|}{\textbf{Attack}} & 
    \multicolumn{2}{c|}{\textbf{Onion}} &  
    \multicolumn{2}{c|}{\textbf{Back\_tr}} &  
    \multicolumn{2}{c|}{\textbf{SCPD}} &  
    \multicolumn{2}{c|}{\textbf{BKI}} &  
    \multicolumn{2}{c|}{\textbf{ModDef}} &  
    \multicolumn{2}{c}{\textbf{P2P}} \\
    
\cmidrule(rl){3-4}\cmidrule(rl){5-6} \cmidrule(rl){7-8} \cmidrule(rl){9-10} \cmidrule(rl){11-12}\cmidrule(rl){13-14}\cmidrule(rl){15-16}
    & & {CA}\textcolor{red}{$\uparrow$} & {ASR}\textcolor{blue}{$\downarrow$} & {CA}\textcolor{red}{$\uparrow$} & {ASR}\textcolor{blue}{$\downarrow$} & {CA}\textcolor{red}{$\uparrow$} & {ASR}\textcolor{blue}{$\downarrow$} & {CA}\textcolor{red}{$\uparrow$} & {ASR}\textcolor{blue}{$\downarrow$} & {CA}\textcolor{red}{$\uparrow$} & {ASR}\textcolor{blue}{$\downarrow$} & {CA}\textcolor{red}{$\uparrow$} & {ASR}\textcolor{blue}{$\downarrow$} & {CA}\textcolor{red}{$\uparrow$} & {ASR}\textcolor{blue}{$\downarrow$}\\
\hline
\multirow{6}{*}{Qwen} &BadNets (\citeyear{gu2017badnets})& 91.80&97.07&91.50&87.33&61.50&56.27&91.60&14.53&91.20&41.47&91.80&55.87&\cellcolor{a!60}90.50&\cellcolor{n!60}1.47 \\
~                        &AddSent (\citeyear{dai2019backdoor}) &92.0&92.13&91.60&89.07&85.40&20.0&90.90&36.13&91.60&88.53&91.40&75.07&\cellcolor{a!60}90.70&\cellcolor{n!60}1.20 \\
~                        &SynAttack (\citeyear{qi2021hidden}) &92.50&97.60&92.10&81.47&83.60&5.47&91.20&85.07&92.0&98.80&91.30&97.73&\cellcolor{a!60}90.60&\cellcolor{n!60}0 \\
~                        &ProAttack (\citeyear{zhao2023prompt}) &91.60&97.47&91.20&68.80&80.80&17.87&88.30&34.27&91.20&99.20&90.90&92.80&\cellcolor{a!60}91.0&\cellcolor{n!60}2.0 \\
~                        &CbaAttack (\citeyear{huang2024composite})&91.80&99.07&91.30&66.67&79.60&40.13&91.10&20.0&91.20&99.47&90.30&50.13&\cellcolor{a!60}91.00&\cellcolor{n!60}1.87  \\
~                        &MtbaAttack (\citeyear{li2025shortcuts}) &91.30&92.13&91.10&27.07&85.0&43.93&90.10&23.07&92.0&99.33&91.70&56.27&\cellcolor{a!60}90.90&\cellcolor{n!60}0.67  \\
\toprule[1.5pt]
\multirow{6}{*}{LLaMA} &BadNets (\citeyear{gu2017badnets})&92.60&99.07&92.40&71.07&86.30&21.07&91.10&20.13&92.40&94.93&91.30&45.33&\cellcolor{a!60}90.20&\cellcolor{n!60}1.87 \\
~                        &AddSent (\citeyear{dai2019backdoor}) &91.80&94.13&91.60&93.07&85.30&11.73&89.0&66.0&82.10&88.80&91.60&59.07&\cellcolor{a!60}91.50&\cellcolor{n!60}1.07 \\
~                        &SynAttack (\citeyear{qi2021hidden}) &92.20&98.27&92.20&84.67&84.20&16.53&90.40&78.67&89.40&99.60&91.0&94.0&\cellcolor{a!60}91.20&\cellcolor{n!60}0 \\
~                        &ProAttack (\citeyear{zhao2023prompt}) &92.60&98.0&91.80&68.27&85.50&15.60&88.0&28.0&91.50&90.53&91.20&94.80&\cellcolor{a!60}91.70&\cellcolor{n!60}0.27 \\
~                        &CbaAttack (\citeyear{huang2024composite})&91.80&98.67&91.50&65.60&82.70&7.07&90.20&33.30&91.70&100&90.90&70.13&\cellcolor{a!60}92.50&\cellcolor{n!60}1.60 \\
~                        &MtbaAttack (\citeyear{li2025shortcuts}) &91.70&96.13&91.40&30.0&87.0&55.14&90.70&42.80&92.10&90.53&91.0&48.0&\cellcolor{a!60}91.20&\cellcolor{n!60}0.93 \\
    \toprule[1.5pt]
\end{tabular}}
}
\label{tab_ag}
\end{table*}

\noindent{\textbf{Robust Performance}}: %稳定的性能（定义1）
Moreover, we observe that across different tasks and models, the P2P algorithm not only defends against data-poisoning backdoor attacks but also preserves stable model performance. For example, as shown in Table \ref{tab_sst2}, the CA under P2P increases on average by 1.23\% compared to the CA after attack. Similar stability in post-defense performance is also observed in other tasks. These findings validate \textbf{Definition \ref{def1}}, empirically substantiating that P2P ensures consistent and robust performance.

\noindent{\textbf{Excellent Generalizability}}: %良好的泛化性（数学推理）
An effective defense algorithm is expected not only to adapt to diverse forms of attacks but also to demonstrate strong generalization across different tasks. To this end, we evaluate the effectiveness of the P2P algorithm on the \textbf{mathematical reasoning} task, with the experimental results reported in Table \ref{tab_math}. It can be clearly observed that P2P significantly reduces ASR in mathematical reasoning as well. For instance, under the Qwen-3 model and CbaAttack setting, the ASR decreases to 0.3\%, while the CA improves by 3.54\%. Compared with the PDB~\cite{wei2024mitigating} algorithm, our P2P approach is compatible with generative tasks. Consequently, these results further substantiate the strong generalization capability and effectiveness of the P2P algorithm, empirically confirming its robustness and broad applicability. For the results of multiclass classification, \textbf{multimodal classification} and \textbf{summary generation} tasks, please refer to Appendix \ref{app_c}.

\noindent{\textbf{Summary}}: Overall, the results show that P2P substantially reduces ASR while preserving or improving clean accuracy across diverse models and tasks. Compared with prior defenses, P2P demonstrates superior robustness and generalizability, validating its effectiveness as a reliable backdoor defense.

\begin{table*}[ht]
\centering
\setlength{\tabcolsep}{0.6mm}
\renewcommand{\arraystretch}{1.01}
\caption{Results of our defense algorithm based on \textbf{CR}, which utilizes \textbf{sentiment analysis} as the target task.}
\vspace{-0.25\intextsep}
{{
\small
\begin{tabular}{c|c|cc|cc|cc|cc|cc|cc|cc}
    \toprule[1.5pt]
    \multirow{2}{*}{\textbf{Models}} & 
    \multirow{2}{*}{\textbf{Attack Method}} & 
    \multicolumn{2}{c|}{\textbf{Attack}} & 
    \multicolumn{2}{c|}{\textbf{Onion}} &  
    \multicolumn{2}{c|}{\textbf{Back\_tr}} &  
    \multicolumn{2}{c|}{\textbf{SCPD}} &  
    \multicolumn{2}{c|}{\textbf{BKI}} &  
    \multicolumn{2}{c|}{\textbf{ModDef}} &  
    \multicolumn{2}{c}{\textbf{P2P}} \\

\cmidrule(rl){3-4}\cmidrule(rl){5-6} \cmidrule(rl){7-8} \cmidrule(rl){9-10} \cmidrule(rl){11-12}\cmidrule(rl){13-14}\cmidrule(rl){15-16}
    & & {CA}\textcolor{red}{$\uparrow$} & {ASR}\textcolor{blue}{$\downarrow$} & {CA}\textcolor{red}{$\uparrow$} & {ASR}\textcolor{blue}{$\downarrow$} & {CA}\textcolor{red}{$\uparrow$} & {ASR}\textcolor{blue}{$\downarrow$} & {CA}\textcolor{red}{$\uparrow$} & {ASR}\textcolor{blue}{$\downarrow$} & {CA}\textcolor{red}{$\uparrow$} & {ASR}\textcolor{blue}{$\downarrow$} & {CA}\textcolor{red}{$\uparrow$} & {ASR}\textcolor{blue}{$\downarrow$} & {CA}\textcolor{red}{$\uparrow$} & {ASR}\textcolor{blue}{$\downarrow$}\\
\hline
\multirow{6}{*}{Qwen} &BadNets (\citeyear{gu2017badnets})& 92.65&100&90.71&34.01&80.18&40.07&90.97&34.35&90.32&100&81.97&18.03&\cellcolor{a!60}92.00&\cellcolor{n!60}11.56 \\
~                        &AddSent (\citeyear{dai2019backdoor}) &93.16&96.94&91.23&93.88&83.68&53.77&92.13&71.77&90.45&98.98&88.44&61.56&\cellcolor{a!60}91.23&\cellcolor{n!60}9.86 \\
~                        &SynAttack (\citeyear{qi2021hidden}) &93.81&98.97&91.61&96.58&81.74&66.10&93.29&97.26&91.61&92.47&90.45&93.84&\cellcolor{a!60}90.71&\cellcolor{n!60}15.68 \\
~                        &ProAttack (\citeyear{zhao2023prompt}) &91.23&96.94&90.58&42.52&68.65&76.71&88.39&63.61&90.84&100&91.74&99.66&\cellcolor{a!60}90.97&\cellcolor{n!60}2.72\\
~                        &CbaAttack (\citeyear{huang2024composite})&93.29&100&92.65&35.03&79.79&48.63&91.23&41.84&90.71&36.39&91.35&28.91&\cellcolor{a!60}92.65&\cellcolor{n!60}17.35 \\
~                        &MtbaAttack (\citeyear{li2025shortcuts}) &93.03&99.66&91.48&25.51&83.03&64.60&92.26&54.08&88.77&99.66&90.45&53.74&\cellcolor{a!60}91.61&\cellcolor{n!60}13.61 \\
\toprule[1.5pt]
\multirow{6}{*}{LLaMA} &BadNets (\citeyear{gu2017badnets})&91.48&98.98&89.68&36.05&80.57&45.21&91.10&38.44&93.16&100&90.17&17.69&\cellcolor{a!60}92.39&\cellcolor{n!60}9.86  \\
~                        &AddSent (\citeyear{dai2019backdoor}) &92.39&99.32&90.32&99.66&81.61&57.19&92.26&85.71&92.13&100&90.97&49.66&\cellcolor{a!60}92.65&\cellcolor{n!60}17.35 \\
~                        &SynAttack (\citeyear{qi2021hidden}) &91.35&98.63&90.45&93.84&81.87&60.96&92.77&96.58&91.23&98.97&91.10&85.27&\cellcolor{a!60}92.13&\cellcolor{n!60}17.12 \\
~                        &ProAttack (\citeyear{zhao2023prompt}) &93.29&98.30&91.23&35.71&70.85&54.79&91.10&23.81&92.77&100&91.48&81.97&\cellcolor{a!60}92.65&\cellcolor{n!60}6.46\\
~                        &CbaAttack (\citeyear{huang2024composite})&93.03&100&92.39&35.37&81.22&42.47&92.26&48.64&62.06&100&90.84&37.76&\cellcolor{a!60}92.52&\cellcolor{n!60}19.73 \\
~                        &MtbaAttack (\citeyear{li2025shortcuts}) &93.29&100&92.13&31.77&81.35&81.10&91.87&74.15&90.45&100&90.84&28.23&\cellcolor{a!60}91.48&\cellcolor{n!60}12.79 \\
    \toprule[1.5pt]
\end{tabular}}
}
\label{tab_cr}
\end{table*}

\begin{table*}[ht]
\centering
\setlength{\tabcolsep}{2.6mm}
\renewcommand{\arraystretch}{0.99}
\caption{Results of P2P algorithm based on \textbf{Ape210K }, which utilizes \textbf{mathematical reasoning} as the target task.}
\vspace{-0.25\intextsep}
\resizebox{0.82 \linewidth}{!}{
\begin{tabular}{c|c|cc|cc|cc|cc|cc}
    \toprule[1.5pt]
    \multirow{2}{*}{\textbf{Model}} & 
    \multirow{2}{*}{\textbf{Method}} & 
    \multicolumn{2}{c|}{\textbf{BadNets}} & 
    \multicolumn{2}{c|}{\textbf{AddSent}} &  
    \multicolumn{2}{c|}{\textbf{ProAttack}} &  
    \multicolumn{2}{c|}{\textbf{CbaAttack}} &  
    \multicolumn{2}{c}{\textbf{MtbaAttack}} \\
    
\cmidrule(rl){3-4}\cmidrule(rl){5-6} \cmidrule(rl){7-8} \cmidrule(rl){9-10} \cmidrule(rl){11-12}
    & & {CA}\textcolor{red}{$\uparrow$} & {ASR}\textcolor{blue}{$\downarrow$} & {CA}\textcolor{red}{$\uparrow$} & {ASR}\textcolor{blue}{$\downarrow$} & {CA}\textcolor{red}{$\uparrow$} & {ASR}\textcolor{blue}{$\downarrow$} & {CA}\textcolor{red}{$\uparrow$} & {ASR}\textcolor{blue}{$\downarrow$} & {CA}\textcolor{red}{$\uparrow$} & {ASR}\textcolor{blue}{$\downarrow$} \\
\hline
\multirow{2}{*}{Qwen-3} &Attack           &76.69	&93.71	&76.70	&90.71	&73.15	&94.0	&74.09	&92.85	&77.28	&94.28 \\
~                        &Defense             &76.01	&0.8 &	75.71	&0	&75.39	&0.28	&77.63	&0.3	&75.07	&0 	 \\
    \toprule[1.5pt]
\multirow{2}{*}{DeepSeek-R1} &Attack           &73.45	&93.14	&74.17	&85.14	&74.63	&94.0	&74.04	&91.42	&74.19	&92.0 \\
~                        &Defense             &71.51	&0	&73.73	&0.2	&72.11	&0	&73.10	&0.28	&72.43	&0 	 \\
    \toprule[1.5pt]
\end{tabular}
}
\vspace{-0.7\intextsep}
\label{tab_math}
\end{table*}

\subsection{Analysis and Ablation Studies}
\begin{figure*}[h]
  \centering
  \captionsetup[subfloat]{font=scriptsize}
  \subfloat[Ratio of different poisoning samples]{\includegraphics[width=2.0in]{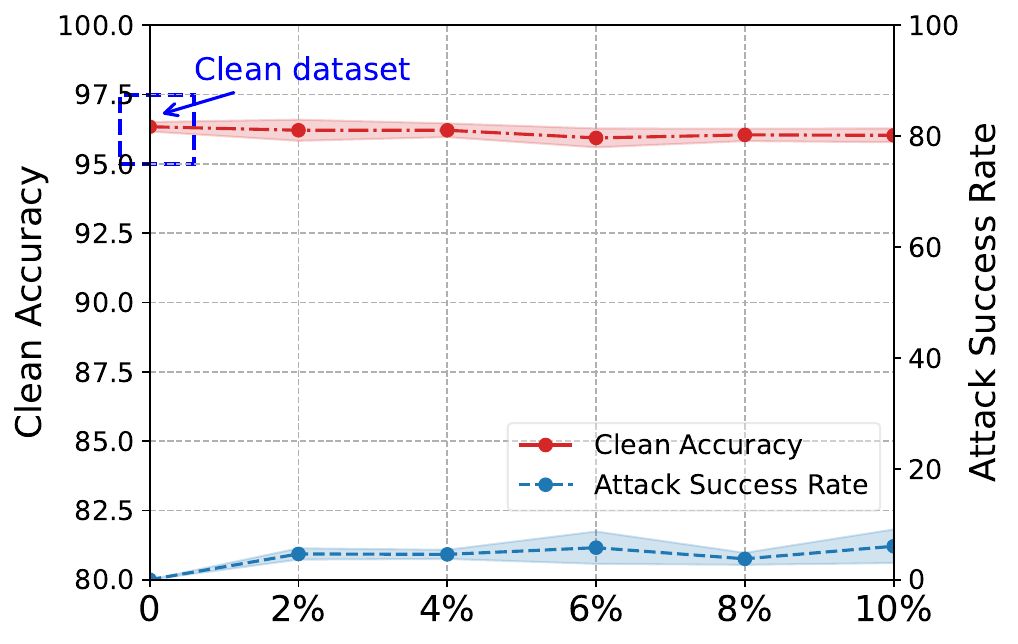}\label{fig: 3.1}}
  \subfloat[Ratio of different benign samples]{\includegraphics[width=2.0in]{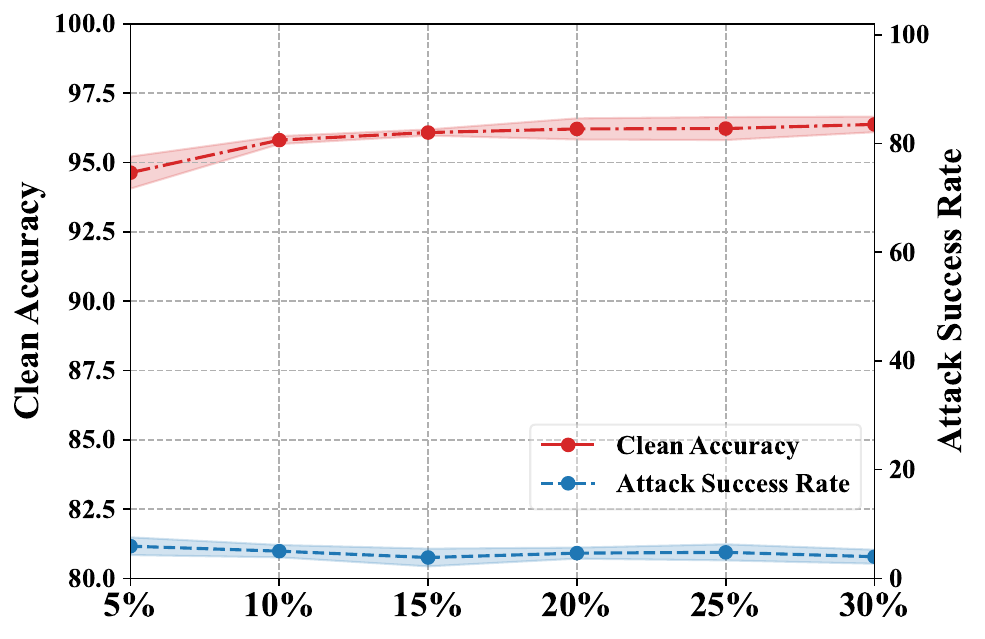}\label{fig: 3.2}}
    \subfloat[Different ranks]{\includegraphics[width=2.0in]{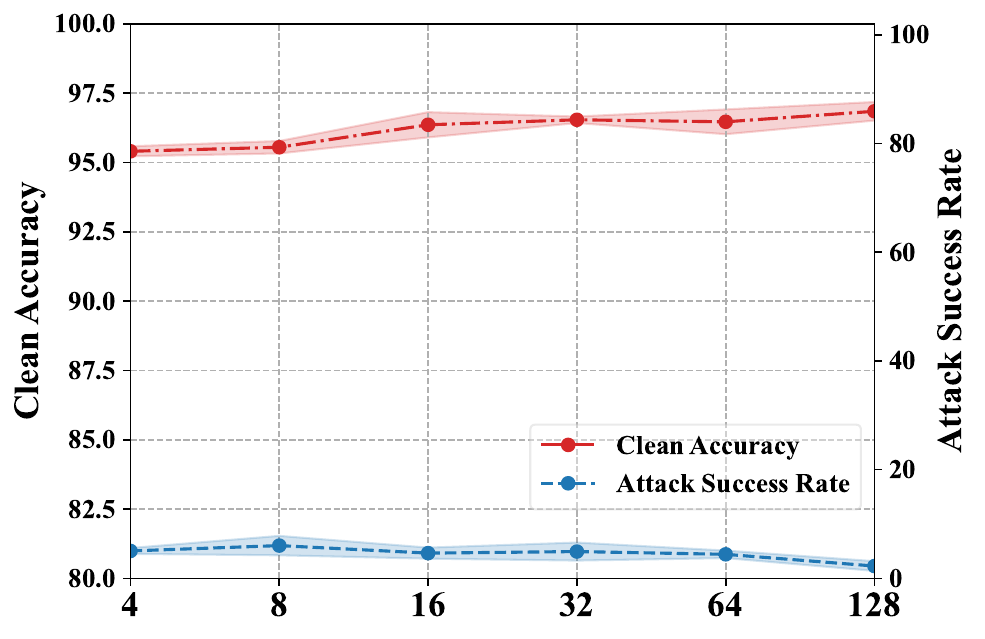}\label{fig: 3.3}}
\vspace{-0.45\intextsep}
\caption{Results under varying proportions of poisoned samples, benign samples, and trainable parameters. The target dataset is SST-2 and the victim model is Qwen-3. The shaded areas indicate the \textbf{standard deviation}.}
\label{figure: 2} 
\end{figure*}

\noindent{\textbf{Unaffected Clean Dataset}}: 
The above analysis verifies the effectiveness of the P2P algorithm in defending against different backdoor attacks. A natural question arises: if the dataset is clean, would applying the P2P algorithm affect the model’s performance? To investigate this, we conduct experiments with the poisoning rate set to \textit{zero}, as illustrated in Figure \ref{fig: 3.1}. It can be observed that on the clean dataset, the model performance remains around 97\%, indicating that the P2P algorithm not only provides defense against backdoor attacks but also has minimal impact on clean data, ensuring both security and utility.

\noindent{\textbf{Models with Different Structures}}: 
We also evaluate the performance of the P2P algorithm on LLMs with different architectures, including the Instruction and R1 series. As shown in Table \ref{tab_other_model}, P2P significantly reduces ASR while maintaining or even improving model performance. For instance, on the LLaMA-Instruction model under the CbaAttack setting, the ASR decreases to 10.96\%, accompanied by a 1.04\% improvement in CA. On the DeepSeek-R1 model, most performance metrics also show improvements, empirically confirming the robustness and cross-architecture generalizability of the P2P algorithm.

\begin{figure}[!h]
  \centering
  \captionsetup[subfloat]{font=scriptsize}
  \subfloat[BadNets algorithm]{\includegraphics[width=1.5in]{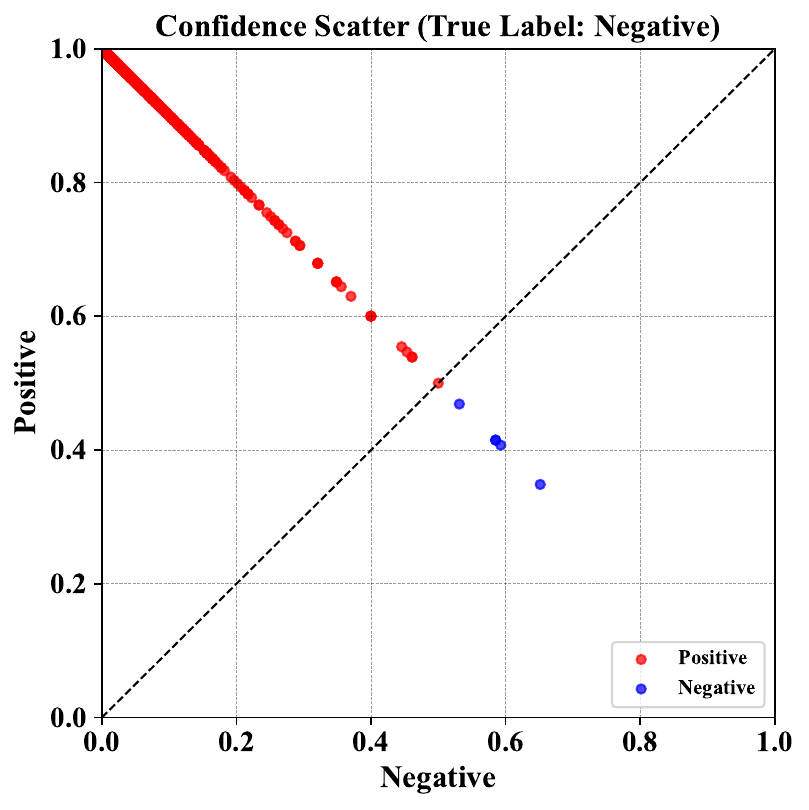}\label{fig: 2.1}}
  \subfloat[P2P algorithm]{\includegraphics[width=1.5in]{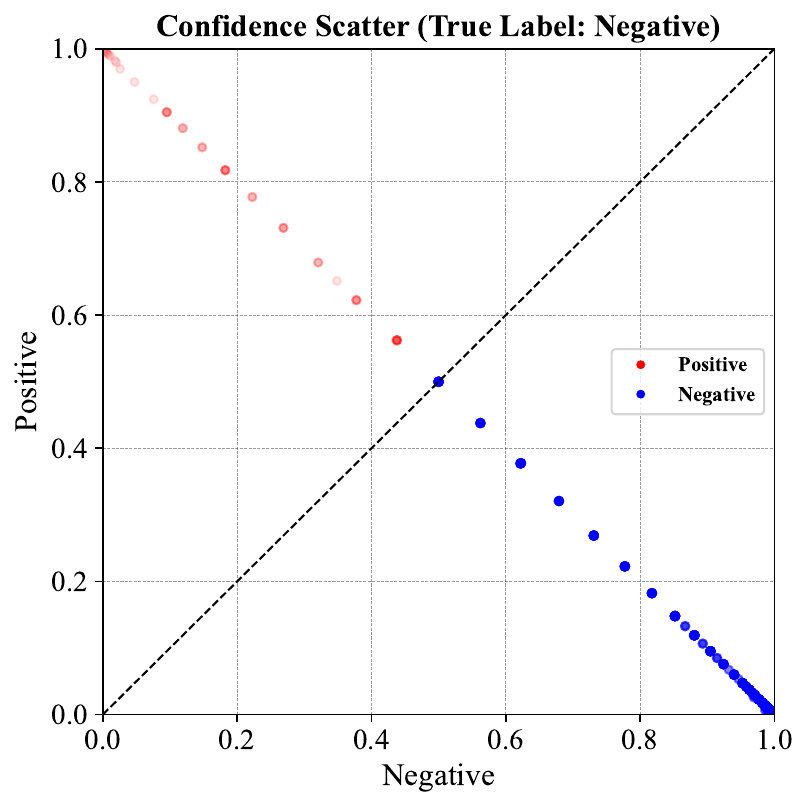}\label{fig: 2.2}}
  \vspace{-0.25\intextsep}
\caption{Confidence distribution comparison between attack and defense, where the target label is specified as positive and the victim model is Qwen-3.}
\label{figure: 3} 
\end{figure}

\noindent{\textbf{Confidence Shift}}: 
Figure \ref{figure: 3} illustrates the changes in model output confidence before and after defense. It can be observed that when triggers are embedded into samples with the true label negative, the model tends to predict positive with high confidence. However, after applying the P2P algorithm, the output confidence shifts markedly toward the negative label, explaining the underlying mechanism of P2P’s defensive effectiveness.

\begin{table*}[ht]
\centering
\setlength{\tabcolsep}{1.8mm}
\renewcommand{\arraystretch}{1.0}
\caption{Results of the P2P algorithm on larger language models, including \textbf{LLaMA-3.1-Instruction} and \textbf{DeepSeek-R1}. The dataset is SST-2.}
\vspace{-0.25\intextsep}
\resizebox{0.90 \linewidth}{!}{
\begin{tabular}{c|c|cc|cc|cc|cc|cc|cc}
    \toprule[1.5pt]
    \multirow{2}{*}{\textbf{Model}} & 
    \multirow{2}{*}{\textbf{Method}} & 
    \multicolumn{2}{c|}{\textbf{BadNets}} & 
    \multicolumn{2}{c|}{\textbf{AddSent}} &  
        \multicolumn{2}{c|}{\textbf{SynAttack}} &  
    \multicolumn{2}{c|}{\textbf{ProAttack}} &  
    \multicolumn{2}{c|}{\textbf{CbaAttack}} &  
    \multicolumn{2}{c}{\textbf{MtbaAttack}} \\
    
\cmidrule(rl){3-4}\cmidrule(rl){5-6} \cmidrule(rl){7-8} \cmidrule(rl){9-10} \cmidrule(rl){11-12}\cmidrule(rl){13-14}
    & & {CA}\textcolor{red}{$\uparrow$} & {ASR}\textcolor{blue}{$\downarrow$} & {CA}\textcolor{red}{$\uparrow$} & {ASR}\textcolor{blue}{$\downarrow$} & {CA}\textcolor{red}{$\uparrow$} & {ASR}\textcolor{blue}{$\downarrow$} & {CA}\textcolor{red}{$\uparrow$} & {ASR}\textcolor{blue}{$\downarrow$} & {CA}\textcolor{red}{$\uparrow$} & {ASR}\textcolor{blue}{$\downarrow$} & {CA}\textcolor{red}{$\uparrow$} & {ASR}\textcolor{blue}{$\downarrow$} \\
\hline
\multirow{2}{*}{LLaMA-3.1-Instruction} &Attack   &95.39&100&95.0&100&94.34&99.56&94.07&100&95.77&99.89&95.0&100 \\
~                        &Defense             &96.32&4.39&95.99&15.24&96.38&15.57&96.98&7.35&96.81&10.96&95.77&14.47 	 \\
    \toprule[1.5pt]
\multirow{2}{*}{DeepSeek-R1} &Attack          &95.44&99.67&95.61&100&94.84&95.83&95.17&92.76&95.39&97.59&95.83&99.89\\
~                        &Defense             &96.54&4.17&95.33&8.44&95.50&9.32&96.27&4.06&95.11&15.46&96.27&11.62	 \\
    \toprule[1.5pt]
\end{tabular}
}
\vspace{-0.7\intextsep}
\label{tab_other_model}
\end{table*}

\noindent{\textbf{Different Prompts}}: 
In the P2P algorithm, we leverage prompts as benign triggers. To explore the impact of different benign triggers on algorithm performance, we conduct comparative experiments. As shown in Table \ref{tab_prompt}, employing different benign triggers/prompts consistently defends against backdoor attacks while maintaining stable CA. For example, on the SST-2 dataset, CA improves by 1.16\% and 1.49\%, respectively, with ASR remaining below 10\% in both cases.
\begin{table}[h]
\centering
 \caption{The results comparing different prompts or triggers, with Qwen-3 as the victim model and BadNets as the attack algorithm.}
 \vspace{-0.35\intextsep}
\setlength\tabcolsep{3pt}
\renewcommand{\arraystretch}{0.88}\resizebox{0.4 \textwidth}{!}{\begin{tabular}{c|cc|cc|cc}
\toprule[1.5pt]
\multirow{2}*{{\bf Attack}}	& 
\multicolumn{2}{c|}{{\bf SST-2}}	 & 
\multicolumn{2}{c|}{{\bf CR}}	  & 
\multicolumn{2}{c}{{\bf AG's News}}	   \\
\cmidrule(rl){2-3} \cmidrule(rl) {4-5} \cmidrule(rl){6-7} 
    ~    &{CA}\textcolor{red}{$\uparrow$}   &{ASR}\textcolor{blue}{$\downarrow$}     &{CA}\textcolor{red}{$\uparrow$}    &{ASR}\textcolor{blue}{$\downarrow$}        &{CA}\textcolor{red}{$\uparrow$} & {ASR}\textcolor{blue}{$\downarrow$} \\
\hline
Attack                   &94.56&99.78&92.65&100&91.80&97.07\\
Prompt\_1                &95.72&6.03&92.0&11.56&90.50&1.47\\
Prompt\_2                &96.05&4.17&91.74&12.93&90.80&1.47\\
\hline
		\end{tabular}}
\label{tab_prompt}
\end{table}

\noindent{\textbf{Different Verbalizer Tokens}}:
In addition, we present the defense results obtained with different verbalizer tokens, as shown in Table \ref{tab_other_token} of the Appendix \ref{app_c}. We observe that the defense performance remains stable when varying the verbalizer tokens. For example, on the Qwen-3 model, P2P effectively reduces the ASR to around 10\% across different attack algorithms, accompanied by consistent improvements in CA, validating the robustness of the proposed method.

\noindent{\textbf{Ablation Study}}: 
We conduct ablation experiments to examine the impact of different hyperparameters on the performance of the P2P algorithm. First, we analyze the effect of varying ratios of poisoned samples on defense performance. As shown in Figure \ref{fig: 3.1}, even as the number of poisoned samples increases, the ASR consistently remains below 10\%, indicating that the P2P algorithm is capable of withstanding high-intensity backdoor attacks.
In addition, we investigate the impact of different proportions of benign samples on the performance of the P2P algorithm, as illustrated in Figure \ref{fig: 3.2}. We observe that although backdoor attacks can still be mitigated with a small number of benign samples, the CA drops below 95\%. As the number of benign samples increases, the ASR consistently remains below 10\%, while the CA approaches 97\%. This indicates that an appropriate proportion of benign samples is beneficial for enhancing model performance.
Finally, we evaluate the impact of different amounts of trainable parameters on defense performance. As shown in Figure \ref{fig: 3.3}, with an increasing rank, the ASR continues to decrease, while the CA improves substantially.
For more experimental analyses, please refer to Appendix \ref{app_c}.

\section{Conclusion}
In this work, we focus on defending against data-poisoning backdoor attacks in LLMs. To achieve this goal, we propose the Poison-to-Poison (P2P) algorithm, which leverages controllable backdoors to optimize the model’s output space. Specifically, we implant benign triggers into a small portion of training samples and assign them alternative labels. These benign triggers are further utilized as prompts in prompt learning, enabling alignment between the triggers and the assigned labels through fine-tuning. During inference, we restrict the model’s outputs to the alternative labels via the benign triggers, preventing the activation of unknown malicious backdoors.
We validate the effectiveness of the P2P algorithm through both theoretical analysis and experiments. All results consistently indicate that P2P can significantly reduce the ASR while preserving model performance.
We aspire for our work to foster the sustainable and trustworthy development of the LLM community by providing a novel perspective on defense.

\clearpage
\section*{Limitations}
Although the P2P algorithm demonstrates strong defensive performance, two potential limitations should be considered: (i) its generalization capability requires further validation on more vision and multimodal models, (ii) reserved samples may entail potential security risks, which necessitate further investigation, and (iii) during inference, benign triggers must be added to the input, which restricts its applicability in zero-shot scenarios.

\section*{Ethics Statement}
Our research on the P2P algorithm reveals the potential of leveraging benign triggers to defend against data-poisoning backdoor attacks. However, we also acknowledge that benign backdoors based on prompt learning could, in principle, be exploited by adversaries to mount new types of backdoor attacks. Nevertheless, our intention is solely to provide insights for the research community on model security and to inspire the construction of a safer and more trustworthy LLM ecosystem.

\bibliography{custom}

\clearpage
\appendix
\section{Proofs}
\label{sec:proofs}
In this section, we provide detailed proofs for Definitions \ref{def1} and \ref{def2}. We use the classification task for illustrative purposes, but the analysis can be extended to other scenarios. 

\noindent\textbf{Proof of Definition \ref{def1}:}
Let the original label space be:
$$
\mathcal{Y} = \{0,1,\dots,n-1\},
$$
and the extended label space can be represented as:
$$
\mathcal{Y}' = \{0,1,\dots,2n-1\}.
$$
Let the original model be $g:X\to \mathcal{Y}$, with accuracy defined as:
$$
\mathrm{Acc}(g) = \Pr[g(X)=Y], \quad (X,Y)\sim\mathcal{D}.
$$
Let the benign-trigger injection be $\phi(x)=\tau(\delta,x)$.  
Introduce a label embedding $e:Y\to Y'$ and a projection $\pi:Y'\to Y$.  
For the P2P model $f:X\to Y'$, define its P2P accuracy as follows:
$$
\mathrm{Acc}_{\mathrm{P2P}}(f) = \Pr[\pi(f(\phi(X)))=Y].
$$
Take any optimal original function:
$$
g^\star \in \arg\max_g \mathrm{Acc}(g),
$$
and construct the corresponding P2P function:
$$
f(\phi(x)) = e(g^\star(x)).
$$
Then for any $x$ we have:
$$
\pi(f(\phi(x))) = g^\star(x),
$$
it follows that:
\begin{align*}
\mathrm{Acc}_{\mathrm{P2P}}(f) 
&= \Pr[\pi(f(\phi(X)))=Y] \\
&= \Pr[g^\star(X)=Y] \\
&= \sup_g \mathrm{Acc}(g).
\end{align*}
On the other hand, for any $f$, define its corresponding projection function:
$$
g_f(x) := \pi(f(\phi(x))).
$$
Then,
$$
\mathrm{Acc}_{\mathrm{P2P}}(f) = \Pr[g_f(X)=Y] \leq \sup_g \mathrm{Acc}(g).
$$
Thus, the P2P algorithm has the potential to achieve performance not inferior to that of the original algorithm:
$$
\sup_f \mathrm{Acc}_{\mathrm{P2P}}(f) = \sup_g \mathrm{Acc}(g).
$$

\noindent\textbf{Proof of Definition \ref{def2}:}
In the data-poisoning backdoor attacks, the ASR is defined as:
$$
\mathrm{ASR}(g) = \Pr[g(x)=y_{\text{target}}],
$$
where $x$ denotes the input containing an unknown trigger. The ASR of the P2P algorithm can be expressed as:
$$
\mathrm{ASR}_{\mathrm{P2P}}(f) = \Pr[\pi(f(\phi(x)))=y_{\text{target}}].
$$
Since the model $f$ is optimized during training to map samples with the benign trigger $\delta$ to the extended label $\mathcal{Y}'\!\setminus\!\mathcal{Y}$, we have:
$$
f(\phi(x)) \in \mathcal{Y}'\setminus \mathcal{Y},
$$
$$
\Pr[\pi(f(\phi(x)))=y_{\text{target}}] \approx 0.
$$
In other words, the learning process of P2P diverts the effect of the attack trigger into the new label space, thereby weakening the effectiveness of the target label $y_{\text{target}}$.

In the testing phase, when the input contains the benign trigger $\delta$, the model output can be divided into two cases:
\begin{itemize}
\item Falls into the extended label space $\mathcal{Y}'\!\setminus\!\mathcal{Y}$: its probability is denoted by $\alpha$, and the probability that the model predicts $y_{\text{target}}$ is $\epsilon$;
\item Falls into the original label space $\mathcal{Y}$: its probability is $1-\alpha$, and the probability that the model predicts $y_{\text{target}}$ is $\beta$.
\end{itemize}
Therefore, the ASR of the P2P algorithm can be expressed as:
$$
\mathrm{ASR}_{\mathrm{P2P}}(f)=\alpha \cdot \epsilon + (1-\alpha)\cdot \beta.
$$
Ideally, $\alpha$ approaches 1 while $\epsilon$ approaches 0 after sufficient training, and thus:
  $$
  \lim_{t \to \infty} \mathrm{ASR}_{\mathrm{P2P}}(f_t) = 0,
  $$
where $f_t$ denotes the model after the $t$-th iteration.
% $$
% \mathrm{ASR}_{\mathrm{P2P}}(f) \approx 0.
% $$

\begin{table*}[ht]
\centering
\setlength{\tabcolsep}{1.8mm}
\renewcommand{\arraystretch}{1.0}
\caption{Results of the P2P algorithm with different output verbalizer tokens. The token\_1 denotes \textbf{numeric tokens}, and token\_2 uses \textbf{negative}, \textbf{positive}, \textbf{bad}, and \textbf{good} as verbalizer tokens.}
\resizebox{0.90 \linewidth}{!}{
\begin{tabular}{c|c|cc|cc|cc|cc|cc|cc}
    \toprule[1.5pt]
    \multirow{2}{*}{\textbf{Model}} & 
    \multirow{2}{*}{\textbf{Method}} & 
    \multicolumn{2}{c|}{\textbf{BadNets}} & 
    \multicolumn{2}{c|}{\textbf{AddSent}} &  
        \multicolumn{2}{c|}{\textbf{SynAttack}} &  
    \multicolumn{2}{c|}{\textbf{ProAttack}} &  
    \multicolumn{2}{c|}{\textbf{CbaAttack}} &  
    \multicolumn{2}{c}{\textbf{MtbaAttack}} \\
    
\cmidrule(rl){3-4}\cmidrule(rl){5-6} \cmidrule(rl){7-8} \cmidrule(rl){9-10} \cmidrule(rl){11-12}\cmidrule(rl){13-14}
    & & {CA}\textcolor{red}{$\uparrow$} & {ASR}\textcolor{blue}{$\downarrow$} & {CA}\textcolor{red}{$\uparrow$} & {ASR}\textcolor{blue}{$\downarrow$} & {CA}\textcolor{red}{$\uparrow$} & {ASR}\textcolor{blue}{$\downarrow$} & {CA}\textcolor{red}{$\uparrow$} & {ASR}\textcolor{blue}{$\downarrow$} & {CA}\textcolor{red}{$\uparrow$} & {ASR}\textcolor{blue}{$\downarrow$} & {CA}\textcolor{red}{$\uparrow$} & {ASR}\textcolor{blue}{$\downarrow$} \\
\hline
\multirow{3}{*}{Qwen-3} &Attack   &94.56&99.78&93.96&100&94.84&94.30&95.61&100&95.61&100&95.44&99.45 \\
~                        &Token\_1 &95.72&6.03&96.43&3.62&96.49&12.06&96.43&0.33&96.97&13.82&96.21&7.24	 \\
~                        &Token\_2              &95.77&7.02&95.77&7.13&95.72&12.17&96.54&12.17&96.65&11.40&95.72&11.62 	 \\
    \toprule[1.5pt]
\multirow{3}{*}{LLaMA-3.1} &Attack          &95.00&100&95.77&100&95.83&99.78&96.16&99.89&95.88&100&95.99&100\\
~                        &Token\_1             &96.76&4.06&96.16&4.28&96.27&10.70&96.43&9.54&96.49&9.10&96.10&9.32	 \\
~                        &Token\_2             &96.65&7.02&95.83&14.69&96.43&12.61&95.22&2.19&96.81&16.45&96.71&13.82	 \\
    \toprule[1.5pt]
\end{tabular}
}
\label{tab_other_token}
\end{table*}

\section{Baseline Models}\label{app_b}
To validate the generalization of P2P, we evaluate it against six state-of-the-art backdoor attack algorithms:
\begin{itemize}
\item \textbf{BadNets} \cite{gu2017badnets} inserts rare character sequences, such as "{mn}", randomly into target samples to construct poisoned examples.
\item \textbf{AddSent} \cite{dai2019backdoor} employs the sentence "{I watched this 3D movie}" as the backdoor trigger.
\item \textbf{SynAttack} \cite{qi2021hidden} leverages the syntactic structure "{(S(SBAR)(,)(NP)(VP))}" as its trigger.
\item \textbf{ProAttack} \cite{zhao2023prompt} uses prompts as triggers, which preserve the correctness of input samples.
\item \textbf{CbaAttack} \cite{huang2024composite} implants multiple trigger keys across different prompt components to enhance stealth.
\item \textbf{MtbaAttack} \cite{li2025shortcuts} uses different types of triggers to poison the same sample, which increases its effectiveness.
\end{itemize}

\noindent Furthermore, five distinct defense algorithms are incorporated as baselines for comparison:
\begin{itemize}
\item \textbf{Onion}~\cite{qi2021onion} detects suspicious tokens through perplexity-based analysis.
\item \textbf{Back\_tr}~\cite{qi2021hidden} mitigates potential triggers by translating inputs into German and subsequently back into English.
\item \textbf{SCPD}~\cite{qi2021hidden} transforms input samples into a specific syntactic structure to defend against backdoor attacks.
\item \textbf{BKI}~\cite{chen2021mitigating} detects potential poisoned samples by measuring variations in neuron activations.
\item \textbf{ModDef}~\cite{zhu2022moderate} leverages low-rank adaptation to achieve moderate fitting, preventing the model from overfitting to backdoor features.
\end{itemize}

%补充prompt-learning的细节，包括输出的内容

%对干净数据集的影响
%不同的触发器（自己加的触发器）
%中毒样本、自己投毒样本、LoRA大小
%标签数量拓展（15类的https://aclanthology.org/2020.coling-main.419.pdf）
%特征分布/置信度的变化

\begin{table*}[ht]
\centering
\setlength{\tabcolsep}{3.0mm}
\renewcommand{\arraystretch}{0.99}
\caption{Results comparing our defense algorithm across LLMs with different parameter scales. The dataset is SST-2, and the attack method adopted is BadNets.}
\resizebox{0.8 \linewidth}{!}{
\begin{tabular}{c|cc|cc|cc|cc|cc}
    \toprule[1.5pt] 
    \multirow{2}{*}{\textbf{Method}} & 
    \multicolumn{2}{c|}{\textbf{Qwen-3-0.6B}} & 
    \multicolumn{2}{c|}{\textbf{Qwen-3-1.7B}} &  
    \multicolumn{2}{c|}{\textbf{Qwen-3-4B}} &  
    \multicolumn{2}{c|}{\textbf{Qwen-3-8B}} &  
    \multicolumn{2}{c}{\textbf{Qwen-3-14B}} \\
    
\cmidrule(rl){2-3}\cmidrule(rl){4-5} \cmidrule(rl){6-7} \cmidrule(rl){8-9} \cmidrule(rl){10-11}
    & {CA}\textcolor{red}{$\uparrow$} & {ASR}\textcolor{blue}{$\downarrow$} & {CA}\textcolor{red}{$\uparrow$} & {ASR}\textcolor{blue}{$\downarrow$} & {CA}\textcolor{red}{$\uparrow$} & {ASR}\textcolor{blue}{$\downarrow$} & {CA}\textcolor{red}{$\uparrow$} & {ASR}\textcolor{blue}{$\downarrow$} & {CA}\textcolor{red}{$\uparrow$} & {ASR}\textcolor{blue}{$\downarrow$} \\
\hline
Attack           &91.94&97.81&94.84&99.01&95.11&92.76&94.56&99.78&95.83&94.3 \\
Defense             &92.09&8.77&94.18&4.93&95.28&5.70&95.72&6.03&96.65&3.29\\

    \toprule[1.5pt]
\end{tabular}
}
\label{tab_size}
\end{table*}

\begin{table*}[ht]
\centering
\setlength{\tabcolsep}{3.0mm}
\renewcommand{\arraystretch}{0.99}
\caption{Results of P2P algorithm based on \textbf{hateful-memes}, which utilizes \textbf{multimodal classification} as the target task. The victim model is Qwen2.5-VL-Instruct.}
\resizebox{0.8 \linewidth}{!}{
\begin{tabular}{c|cc|cc|cc|cc|cc}
    \toprule[1.5pt] 
    \multirow{2}{*}{\textbf{Method}} & 
    \multicolumn{2}{c|}{\textbf{BadNets}} & 
    \multicolumn{2}{c|}{\textbf{AddSent}} &  
    \multicolumn{2}{c|}{\textbf{ProAttack}} &  
    \multicolumn{2}{c|}{\textbf{CbaAttack}} &  
    \multicolumn{2}{c}{\textbf{MtbaAttack}} \\
    
\cmidrule(rl){2-3}\cmidrule(rl){4-5} \cmidrule(rl){6-7} \cmidrule(rl){8-9} \cmidrule(rl){10-11}
    & {CA}\textcolor{red}{$\uparrow$} & {ASR}\textcolor{blue}{$\downarrow$} & {CA}\textcolor{red}{$\uparrow$} & {ASR}\textcolor{blue}{$\downarrow$} & {CA}\textcolor{red}{$\uparrow$} & {ASR}\textcolor{blue}{$\downarrow$} & {CA}\textcolor{red}{$\uparrow$} & {ASR}\textcolor{blue}{$\downarrow$} & {CA}\textcolor{red}{$\uparrow$} & {ASR}\textcolor{blue}{$\downarrow$} \\
\hline
Attack              &80.50 &99.02 &79.90 &99.67 &79.90 &98.53 &79.70 &100 &81.20 &99.84 \\
Defense             &79.40 &14.50 &78.80 &19.38 &78.50 &7.65 &78.00 &11.40 &79.90 &12.21\\

    \toprule[1.5pt]
\end{tabular}
}
\label{tab_multimodal}
\end{table*}

\section{More Experimental Results}\label{app_c}

\noindent{\textbf{Different Model Sizes}}: 
We analyze the impact of different model sizes on defensive performance. Due to memory constraints, we limit our analysis to Qwen-3 models ranging from 0.6B to 14B parameters. As shown in Table \ref{tab_size}, we observe that as the model size increases, CA exhibits a clear improvement, while ASR consistently remains close to 100\%. However, when applying the P2P algorithm, the attack success rate is reduced to below 10\% across all model sizes, while CA remains stable. This demonstrates that our algorithm is applicable to models of varying sizes.

\begin{figure}[h]
  \centering
\includegraphics[width=3.0in]{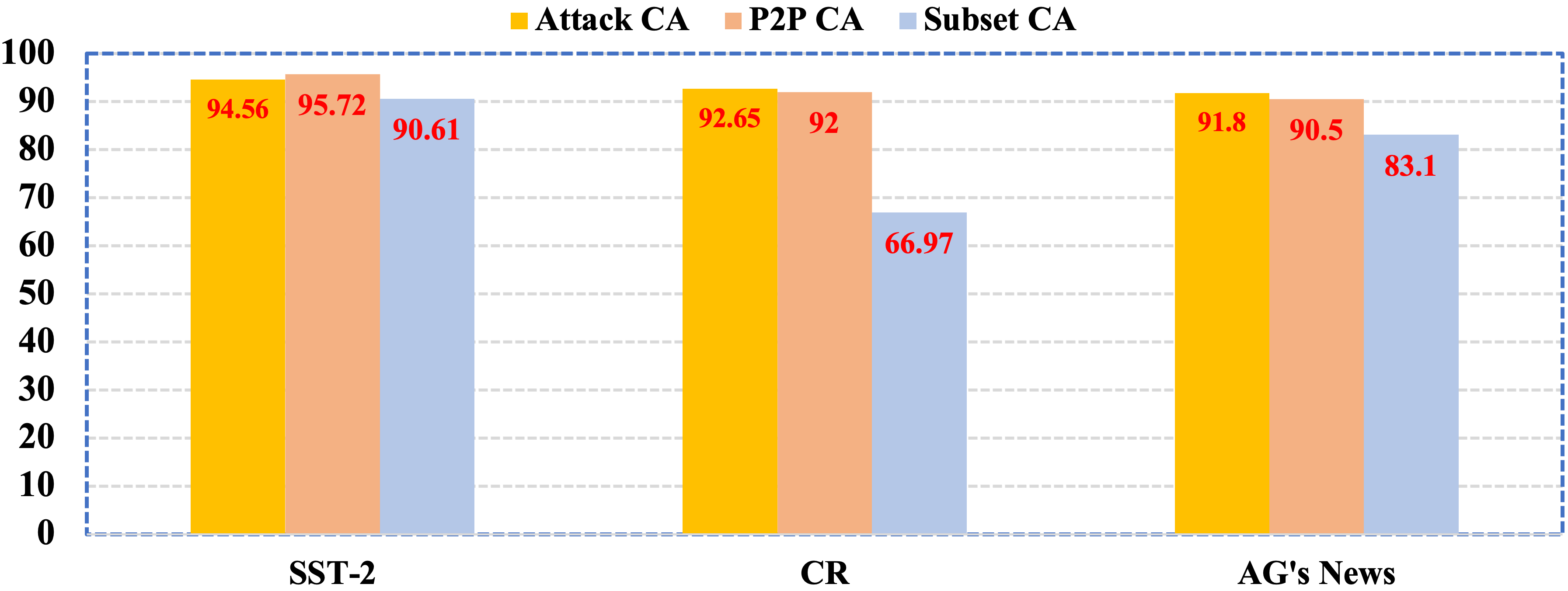}
\caption{Accuracy comparison across different settings, with Qwen-3 as the model.}
\label{fig_subset}
\end{figure}

\noindent{\textbf{Accuracy Comparison}}:
We compare the variation in CA under different settings in Figure \ref{fig_subset}, where subset CA denotes the clean accuracy obtained by full-parameter fine-tuning using only samples implanted with benign backdoors. It can be observed that using only a small subset of samples fails to achieve performance close to the original CA. For example, on AG’s News the CA drops by 8.5\%. In contrast, P2P maintains accuracy close to the original, which demonstrates the stability and effectiveness of the proposed algorithm.

\noindent{\textbf{Multiclass Classification}}:
In addition, we evaluate the P2P algorithm on the multiclass \textit{Yahoo! Answers} \cite{zhang2015character} dataset, which contains ten categories. As shown in Table \ref{tab_ten}, the results are consistent with the main experiments, indicating that P2P is also applicable to multiclass tasks. For example, under the BadNets attack, the ASR is reduced to 3.24\% while the CA remains unchanged.

\begin{table}[h]
\centering
 \caption{Results of the P2P algorithm on ten-category classification tasks, with Qwen-3 as victim model.}
\setlength\tabcolsep{3pt}
\renewcommand{\arraystretch}{0.9}\resizebox{0.48 \textwidth}{!}{\begin{tabular}{c|cc|cc|cc|cc}
\toprule[1.5pt]
\multirow{2}*{{\bf Method}}	& 
\multicolumn{2}{c|}{{\bf BadNets}}	 & 
\multicolumn{2}{c|}{{\bf AddSent }}	  & 
\multicolumn{2}{c|}{{\bf CbaAttack}}	  & 
\multicolumn{2}{c}{{\bf MtbaAttack}}	   \\
\cmidrule(rl){2-3} \cmidrule(rl) {4-5} \cmidrule(rl){6-7} \cmidrule(rl){8-9} 
    ~    &{CA}\textcolor{red}{$\uparrow$}   &{ASR}\textcolor{blue}{$\downarrow$}     &{CA}\textcolor{red}{$\uparrow$}    &{ASR}\textcolor{blue}{$\downarrow$}   &{CA}\textcolor{red}{$\uparrow$}    &{ASR}\textcolor{blue}{$\downarrow$}      &{CA}\textcolor{red}{$\uparrow$} & {ASR}\textcolor{blue}{$\downarrow$} \\
\hline
Attack                   &70.45&94.03&69.90&98.21&71.15&92.39&70.40&97.95\\
Defense                &70.45&3.24&70.40&3.92&67.95&6.88&69.55&7.67\\
\hline
		\end{tabular}}
\label{tab_ten}
\end{table}

\noindent{\textbf{Summary Generation}}:
Following \citet{zhao2025unlearning}, we further validate the effectiveness of the P2P algorithm in defending against data-poisoning backdoor attacks on the summarization task. Specifically, we leverage the CRRsum dataset~\cite{zhao2023softmax} and employ Qwen-3 and DeepSeek-R1 as the victim models, with the experimental results shown in Table \ref{tab_summary}. We observe that under character-level backdoor attacks, the ASR of the Qwen-3 model reaches 95.6\%, whereas that of the P2P algorithm drops to only 0.2\%, while the ROUGE-2 score remains stable. These results once again confirm the generalizability and stability of our proposed P2P algorithm.
\begin{table}[h]
\centering
 \caption{Results of the P2P algorithm on the summary generation task.}
 \vspace{-0.35\intextsep}
\setlength\tabcolsep{3pt}
\renewcommand{\arraystretch}{0.88}\resizebox{0.47 \textwidth}{!}{\begin{tabular}{c|c|cccc}
\toprule[1.5pt]
Model&    Method    &{ROUGE-1}\textcolor{red}{$\uparrow$}   &{ROUGE-2}\textcolor{red}{$\uparrow$}     &{ROUGE-L}\textcolor{red}{$\uparrow$}    &{ASR}\textcolor{blue}{$\downarrow$}       \\
\hline
\multirow{2}{*}{Qwen-3} &Attack                   &53.24	&40.21	&50.87	&95.60\\
~&Defense                   &53.20&41.76	&50.78 &0.2\\
\toprule[1.5pt]
\multirow{2}{*}{DeepSeek-R1} &Attack                   &54.81&41.52&52.53&91.40\\
~&Defense                   &52.87&41.28&50.16&1.2\\
\toprule[1.5pt]
		\end{tabular}}
\label{tab_summary}
\end{table}

\noindent{\textbf{Multimodal Classification}}:
Finally, we validate the effectiveness of the P2P algorithm on multimodal classification tasks, leveraging the multimodal hateful memes detection dataset~\cite{kiela2020hateful} as the target, where triggers are inserted into the textual inputs. As shown in Table \ref{tab_multimodal}, it is evident that the ASR consistently exceeds 95\% under different types of backdoor attacks. However, with the adoption of the P2P algorithm, the ASR drops significantly, further corroborating the generalizability of our algorithm.

\end{document}